\documentstyle[12pt]{article}

\def\thebibliography#1{ {\bf References}
\list
 {[\arabic{enumi}]}{\settowidth\labelwidth{[#1]}\leftmargin\labelwidth
 \advance\leftmargin\labelsep
 \usecounter{enumi}}
 \def\newblock{\hskip .11em plus .33em minus .07em}
 \sloppy\clubpenalty4000\widowpenalty4000
 \sfcode`\.=1000\relax}

\begin{document}
 \thispagestyle{empty}
 \begin{flushright}
 {BERGEN--1995-14}\\
 {MZ-TH--95-26}\\
 {hep-ph/9511261}\\
 {November 1995}\\
\end{flushright}
 \vspace*{3cm}
 \begin{center}
 {\bf \large
 Tensor reduction of two-loop vacuum diagrams\\[2mm]
 and projectors for expanding three-point functions}
 \end{center}
 \vspace{1cm}
 \begin{center}
 A.~I.~Davydychev$^{a,b,}$\footnote{davyd@vsfys1.fi.uib.no}
 \ \ and \ \
 J.~B.~Tausk$^{c,}$\footnote{tausk@vipmzw.physik.uni-mainz.de}\\
 \vspace{1cm}
$^{a}${\em
 Department of Physics, University of Bergen, \\
      All\'{e}gaten 55, N-5007 Bergen, Norway}
\\
\vspace{.3cm}
$^{b}${\em
 Institute for Nuclear Physics,
 Moscow State University, \\
 119899, Moscow, Russia}
\\
\vspace{.3cm}
$^{c}${\em
Institut f\"ur Physik,
  Johannes Gutenberg Universit\"at, \\
Staudinger Weg 7, D-55099 Mainz, Germany}
\end{center}
 \hspace{3in}
 \begin{abstract}
Explicit general formulae for the tensor reduction
of two-loop massive vacuum diagrams are presented.
The problem of calculating the corresponding coefficients
is shown to be equivalent to the problem of constructing
differential operators (projectors) extracting the
coefficients of the momentum expansion of massive scalar
three-point functions (with any number of loops),
so the solution to the latter problem is also given.
 \end{abstract}

\newpage

\vspace*{3mm}


{\bf 1. Introduction}

Tensor reduction of Feynman integrals containing loop momenta with
uncontracted Lorentz indices in the numerator is very important for various
realistic calculations in the Standard Model (and beyond). For one-loop
diagrams with different numbers of external lines, several approaches and
algorithms were developed \cite{one-tensor,ibp}. For two-loop self-energy
diagrams, the problem was considered e.g. in ref. \cite{two-tensor}, whilst
the three-point two-loop case is more complicated and requires results for
the integrals with irreducible numerators \cite{UD3-4}.

The problem of finding general algorithms for the calculation of two-loop
vacuum diagrams is interesting because these diagrams occur in many
important physical problems. We can mention, for example, calculation of
the two-loop effective potential in the Standard Model \cite{FJ+J} and
two-loop contributions to the $\rho$-parameter \cite{Bij,rho}. Furthermore,
consideration of many other problems requires constructing efficient
algorithms for the momentum expansion of two-loop massive diagrams with two
\cite{DT}, three \cite{FT}, or more external legs. The coefficients of such
expansions involve massive vacuum diagrams with tensor numerators. A
procedure for reducing the tensors to scalar numerators enables us to
calculate these coefficients analytically, since general results (including
the relevant terms of the $\varepsilon$-expansion) are known for the
corresponding scalar diagrams \cite{FJ+J,Bij,DT,H+S+M}. In the two-point
case, the reduction is relatively simple because, as all indices are
ultimately contracted with the same external momentum, only completely
symmetric tensors need to be considered. However, starting from the three
point-case, when there are two or more independent external momenta to
contract with, a complete tensor decomposition is required.

The answer is usually easy to obtain when one has only a small number of
integration momenta in the numerator. The situation becomes more
complicated when one needs to calculate the higher-order coefficients of
the expansion. This is the case when we are approaching the physical
singularity (threshold), where the expansion still works but the parameter
we expand in is already not so small. Moreover, in the approach \cite{FT}
(which involves conformal mapping and Pad\'e approximations), in order to
get accurate numerical results beyond the threshold(s), one needs to
calculate expansion coefficients of very high order, and therefore tensor
integrals of very high rank. These are some of the reasons why general
algorithms for the tensor decomposition of two-loop vacuum diagrams are
important.

We would like to mention recent progress in this direction. In
ref.~\cite{Ch-AI93}, a closed expression, in terms of gamma functions, was
given for arbitrary tensor vacuum integrals in which two of the masses are
equal and the third is zero. Furthermore, it was noted that the algorithm
used in that paper for reducing the tensors to scalars can be extended to
more general cases. Some relevant formulae for diagrams with arbitrary
masses, where the numerator is contracted with one or two external vectors,
can be found in refs. \cite{DST,FT2,DT-Pisa}. Uncontracted tensor integrals
were considered in refs.~\cite{DT-Pisa,Tarasov-Pisa}. In \cite{DT-Pisa}, a
general expression for the matrix inverse to one containing the
coefficients required for the tensor decomposition was constructed, while
in \cite{Tarasov-Pisa} a version of the direct formula was presented. While
solving the problem in principle, these results \cite{DT-Pisa,Tarasov-Pisa}
could not completely satisfy us because they were not explicitly symmetric
and the coefficients required for the decomposition were rather cumbersome,
so that their use in computer programs was less efficient than recursive
calculation of the corresponding coefficients. In this paper, we present a
simpler general solution to this problem.

Our results also provide a complete and explicit answer to a different
question, which is related to an alternative approach \cite{FT} to the
expansion of scalar three-point diagrams. Instead of Taylor-expanding all
propagators, one applies certain differential operators to the diagram and
then sets the external momenta to zero, in order to obtain the Taylor
coefficients of the diagram as a whole. In this way, one does not need to
evaluate any tensor integrals, but is faced with the problem of finding the
correct differential operators to use. We shall show that this problem is,
in fact, equivalent to the tensor decomposition of two-loop vacuum
integrals, which enables us to give a closed expression for the operator
that projects out the Taylor coefficient of any given order in the three
external kinematical invariants.

The remainder of this paper is organized as follows. In section 2 the
notation is introduced and the connection between two-loop tensor reduction
and projectors for three-point functions is established. The solutions to
both problems involve a set of universal coefficients, which are calculated
in section 3. A summary and a discussion of the results are contained in
section 4.

\vspace{3mm}

{\bf 2. Definitions}

In this paper, we shall use the following notation:
\begin{equation}
\label{defI}
I\left[ \mbox{something} \right]
\equiv
\int\int \mbox{d}^n p \; \mbox{d}^n q \;
\left\{ \mbox{something} \right\} \;
F\left(p^2, q^2, (pq)\right) ,
\end{equation}
and we are interested in expressing the tensor integrals
\begin{equation}
\label{I}
I\left[ p_{\mu_1} \ldots p_{\mu_{N_1}} \;
         q_{\sigma_1} \ldots q_{\sigma_{N_2}} \right]
\end{equation}
in terms of scalar integrals. In eq.~(\ref{defI}), $n$ is
the space-time dimension \cite{dimreg} and
$F\!\!\left(p^2,q^2,(pq)\right)$ is an arbitrary scalar function
depending on Lorentz invariants of the loop momenta $p$ and $q$.
Usually, it is a product of (powers of) propagators,
\begin{equation}
\label{denom}
\left( p^2\!-\!m_1^2 \right)^{-\nu_1}
\left( q^2\!-\!m_2^2 \right)^{-\nu_2}
\left( (p\!-\!q)^2\!-\!m_3^2 \right)^{-\nu_3} \; ,
\end{equation}
times a polynomial in $p^2$, $q^2$ and $(pq)$, but the
concrete form of this function will not be important for our discussion.

Because $F$ is an even function of $p$ and $q$, it is clear that the
integral (\ref{I}) vanishes when $N_1+N_2$ is odd. When $N_1+N_2$ is even
(from now on, we shall always assume this), since we have no external
momenta, the result must be a Lorentz invariant tensor made out of metric
tensors, and it must be symmetric in the two subsets of indices,
$\{\mu_1,\ldots,\mu_{N_1}\}$ and $\{\sigma_1,\ldots,\sigma_{N_2}\}$. A
basis of tensor structures with these properties can be described as
follows.

Each structure is characterized by three integers, $j_1$, $j_2$ and $j_3$,
such that $2 j_1 + j_3 = N_1$ and  $2 j_2 + j_3 = N_2$. It can be
constructed by taking a product of $j_1$ metric tensors $g_{\mu_i \mu_k}$,
$j_2$ tensors $g_{\sigma_i \sigma_k}$ and $j_3$ tensors $g_{\mu_i
\sigma_k}$, and then symmetrizing it in the $\mu$'s and $\sigma$'s by
taking the sum of all {\em distinct} products of metric tensors we get from
it through permutations of the $\mu$'s and of the $\sigma$'s:
\begin{eqnarray}
\label{j1j2j3}
\lefteqn{
\left\{ j_1, j_2, j_3 \right\}_{
\mu_1 \ldots \mu_{N_1} \sigma_1 \ldots \sigma_{N_2}}
} && \nonumber \\
 & \equiv &
g_{\mu_1 \mu_2} \ldots g_{\mu_{2j_1 - 1} \mu_{2j_1}}
g_{\sigma_1 \sigma_2} \ldots g_{\sigma_{2j_2 - 1} \sigma_{2j_2}}
g_{\mu_{2j_1 + 1} \sigma_{2j_2 + 1}} \ldots
g_{\mu_{2j_1 + j_3} \sigma_{2j_2 + j_3}}
\nonumber \\
 & + & \mbox{permutations} \; .
\end{eqnarray}
The number of terms on the r.h.s.\ is
\begin{equation}
\label{c}
t_{j_1 j_2 j_3} = \frac{{N_1}! \; {N_2}!}
                       {2^{j_1+j_2} \; {j_1}! \; {j_2}! \; {j_3}!} \; .
\end{equation}
Equivalent representations of the tensor structures (\ref{j1j2j3})
are
\begin{eqnarray}
\label{jgen}
\left\{ j_1, j_2, j_3 \right\}
&=& \frac{t_{j_1 j_2 j_3}}{N_1! N_2!} \;
\frac{ \partial^{N_1} }
 {\partial k_1^{\mu_1} \ldots \partial k_1^{\mu_{N_1}} } \;
\frac{ \partial^{N_2} }
 {\partial k_2^{\sigma_1} \ldots \partial k_2^{\sigma_{N_2}} } \;
(k_1^2)^{j_1} (k_2^2)^{j_2} (k_1 k_2)^{j_3}
\nonumber \\
&=& \frac{t_{j_1 j_2 j_3}}{N_1! N_2!} \;
\Box_{11}^{j_1} \, \Box_{22}^{j_2} \, \Box_{12}^{j_3} \;
{k_1}_{\mu_1} \ldots {k_1}_{\mu_{N_1}}
{k_2}_{\sigma_1} \ldots {k_2}_{\sigma_{N_2}} \, ,
\end{eqnarray}
where we have adopted the notation of \cite{FT} for
d'Alembertian operators,
\begin{equation}
\Box_{ij} = g^{\mu \nu}
\frac{\partial^2}{\partial k_i^{\mu} \partial k_j^{\nu}} \; .
\end{equation}
For simplicity, we have suppressed the Lorentz indices on the
l.h.s.\ of (\ref{jgen}).
We note that in \cite{ibp,Ch-AI93}\footnote{
Here, we would like to mention that eq.~(B.4) of \cite{BDST}
is equivalent to the result presented in \cite{ibp}, if one explicitly
calculates the result of applying the d'Alembertians.
}, d'Alembertians were used in a similar way to generate the required
tensor structures.

At given $N_1$ and $N_2$, the number of independent tensor structures
(\ref{j1j2j3}) is
\begin{equation}
\label{T}
T(N_1, N_2) = 1 + \min\left( [N_1/2], [N_2/2] \right) ,
\end{equation}
where $[N_i/2]$ is the integer part of $N_i/2$, and each of these
structures is already determined by one of the $j$'s, e.g., $j_3$. Because
we shall often need to sum over the set of all tensor structures, we
introduce the following notation:
\begin{equation}
\label{jsums}
\sum_{\{j\}} \hspace{5mm} \equiv
\sum_{\begin{array}{c} {}^{j_1, j_2, j_3}\\[-3mm]
                       {}_{2j_1+j_3=N_1}\\[-2mm]
                       {}_{2j_2+j_3=N_2}
      \end{array}} ,
\end{equation}
emphasizing that this is actually just a {\em single} sum.

Now, we return to the integral (\ref{I}) and write it as a
linear combination
\begin{eqnarray}
\label{decomp1}
I\left[ p_{\mu_1} \ldots p_{\mu_{N_1}} \;
         q_{\sigma_1} \ldots q_{\sigma_{N_2}} \right]
= \sum_{\{j\}}
\left\{ j_1, j_2, j_3 \right\} \; I_{j_1 j_2 j_3} \; ,
\end {eqnarray}
with some scalar coefficients $I_{j_1 j_2 j_3}$ which are to be
determined. By contracting (\ref{decomp1}) with each of the structures
(\ref{j1j2j3}) we obtain the following system of $T(N_1,N_2)$ linear
equations for the $I_{j_1 j_2 j_3}$'s:
\begin{equation}
\label{Isys}
t_{j_1 j_2 j_3} \;
I\left[ (p^2)^{j_1} (q^2)^{j_2} (pq)^{j_3} \right]
 = \sum_{\{j'\}}
\chi_{j_1 j_2 j_3; j'_1 j'_2 j'_3}
 I_{j'_1 j'_2 j'_3} \; ,
\end{equation}
where
\begin{equation}
\label{chi}
\chi_{j_1 j_2 j_3; j'_1 j'_2 j'_3}
=
 \left\{j_1,j_2,j_3\right\}_{
\mu_1 \ldots \mu_{N_1} \sigma_1\ldots \sigma_{N_2}}
 \left\{ j'_1, j'_2, j'_3 \right\}^{
\mu_1 \ldots \mu_{N_1} \sigma_1 \ldots \sigma_{N_2}}  \; .
\end{equation}
It is useful to think of the coefficients
$\chi_{j_1 j_2 j_3; j'_1 j'_2 j'_3}$ as the elements of a
$T(N_1,N_2)\times T(N_1,N_2)$ ``contraction matrix'', whose rows and
columns are labelled by the triplets $(j_1,j_2,j_3)$ and
$(j'_1,j'_2,j'_3)$, respectively. We shall denote the inverse
of this matrix by $\phi$, i.e.
\begin{equation}
\sum_{\{j''\}}
\chi_{j_1 j_2 j_3; j''_1 j''_2 j''_3}
\phi_{j''_1 j''_2 j''_3; j'_1 j'_2 j'_3}
 = \delta_{j_1 j_2 j_3; j'_1 j'_2 j'_3}
\equiv
\delta_{j_1 j'_1} \delta_{j_2 j'_2} \delta_{j_3 j'_3} \, .
\end{equation}
Since $\chi$ is a symmetric matrix, $\phi$ is also symmetric:
\begin{equation}
\label{sym:jj'}
   \phi_{j_1 j_2 j_3; j'_1 j'_2 j'_3}
=  \phi_{j'_1 j'_2 j'_3; j_1 j_2 j_3} \; .
\end{equation}
Using $\phi$ to solve the system of equations (\ref{Isys}), we
can re-write eq.~(\ref{decomp1}) as
\begin{equation}
\label{decomp2}
I\left[ p_{\mu_1} \ldots p_{\mu_{N_1}} \;
         q_{\sigma_1} \ldots q_{\sigma_{N_2}} \right]
=\sum_{\{j\}}
 \left\{ j_1, j_2, j_3 \right\}
 \sum_{\{j'\}}
 \phi_{j_1 j_2 j_3; j'_1 j'_2 j'_3} \; t_{j'_1 j'_2 j'_3} \;
I\left[ (p^2)^{j'_1} (q^2)^{j'_2} (pq)^{j'_3} \right]  .
\end{equation}
Thus, we have reduced the tensor integrals (\ref{I}) to
a combination of scalar integrals (carrying the same
total powers of momenta $p$ and $q$) multiplied by the
tensor structures $\left\{ j_1, j_2, j_3 \right\}$.
In the next section, we shall derive explicit
expressions for both $\phi$ and $\chi$.

Here, we address the second question posed in the introduction.
Consider a massive, scalar, three-point Feynman diagram depending
on two independent external momenta $k_1$ and $k_2$, and suppose
we are interested in the coefficients of its Taylor
expansion,
\begin{equation}
\label{eq:3point}
C(k_1,k_2) =
\sum_{j_1,j_2,j_3} c_{j_1 j_2 j_3}
(k_1^2)^{j_1} (k_2^2)^{j_2} (k_1 k_2)^{j_3}
\; .
\end{equation}
Our goal is to construct a set of projection operators that give us the
coefficients $c_{j_1 j_2 j_3}$, using derivatives with respect to
the {\em vectors} $k_1$ and $k_2$. This will allow us to apply them to
the integrand of the Feynman integral through which $C(k_1,k_2)$ is
defined.

Following \cite{FT}, let us apply a product of d'Alembertians to both sides
of (\ref{eq:3point}), and then set $k_1=k_2=0$. This gives
\begin{eqnarray}
\label{boxA}
\left.
\Box_{11}^{j_1} \, \Box_{22}^{j_2} \, \Box_{12}^{j_3} \;
C(k_1,k_2)
\right|_{k_1=k_2=0}
&=& \sum_{\{j'\}}
 c_{j'_1 j'_2 j'_3}
\Box_{11}^{j_1} \, \Box_{22}^{j_2} \, \Box_{12}^{j_3} \;
(k_1^2)^{j'_1} (k_2^2)^{j'_2} (k_1 k_2)^{j'_3}
\nonumber \\
&=& N_1! N_2! \sum_{\{j'\}}
 \; \frac{\chi_{j_1 j_2 j_3; j'_1 j'_2 j'_3}}
                  {t_{j_1 j_2 j_3} t_{j'_1 j'_2 j'_3}} \;
c_{j'_1 j'_2 j'_3} \; ,
\end{eqnarray}
where $N_1=2j_1+j_3$ and $N_2=2j_2+j_3$. The second line
follows from eqs. (\ref{jgen}) and (\ref{chi}).
Solving the system of equations (\ref{boxA}) gives us the
operators we are looking for:
\begin{equation}
\label{projectors}
c_{j_1 j_2 j_3} =
\frac{ t_{j_1 j_2 j_3} }
     { N_1! N_2! }
 \sum_{\{j'\}}
\phi_{j_1 j_2 j_3; j'_1 j'_2 j'_3} \; t_{j'_1 j'_2 j'_3}
\left.
\Box_{11}^{j'_1} \, \Box_{22}^{j'_2} \, \Box_{12}^{j'_3} \;
C(k_1,k_2)
\right|_{k_1=k_2=0}
\, ,
\end{equation}
where the coefficients $\phi_{j_1 j_2 j_3; j'_1 j'_2 j'_3}$
are the same as the ones needed in the tensor reduction formula
(\ref{decomp2})! For the coefficients $c_{0 0 j_3}$ and $c_{j_1 0 j_3}$,
this result, combined with eq.~(\ref{phia023}) presented below, coincides
with the expressions given in \cite{FT}.

Another way to see the connection between (\ref{decomp2}) and
(\ref{projectors}) is by taking the scalar function $C(k_1,k_2)$ in
the latter to be
$I\left[ {\left(k_1p\right)}^{N_1}{\left(k_2q\right)}^{N_2} \right]$,
and noting that
\begin{equation}
I\left[ p_{\mu_1} \ldots p_{\mu_{N_1}} \,
         q_{\sigma_1} \ldots q_{\sigma_{N_2}} \right]
= \frac{1}{N_1! N_2!} \;
\frac{ \partial^{N_1} }
 {\partial k_1^{\mu_1} \ldots \partial k_1^{\mu_{N_1}} } \;
\frac{ \partial^{N_2} }
 {\partial k_2^{\sigma_1} \ldots \partial k_2^{\sigma_{N_2}} } \;
I\left[ {\left(k_1p\right)}^{N_1}
        {\left(k_2q\right)}^{N_2} \right] \, .
\end{equation}
{}From this point of view, it is also clear that the connection can
immediately be generalized: the $\phi$'s occurring in the tensor reduction
of $L$-loop vacuum diagrams are the same as in the projectors for the
Taylor coefficients of $(L+1)$-point scalar diagrams.

\vspace{3mm}

{\bf 3. Results}

We calculate the coefficients $\phi_{j_1 j_2 j_3; j'_1 j'_2 j'_3}$ by
an inductive method, based on recurrence relations,
which enables us to express them in a compact and symmetric way.
Afterwards, we shall obtain $\chi$ by inverting $\phi$.

We start from the following simple formulae for contracting two
of the Lorentz indices of the tensor structures (\ref{j1j2j3}), which can be
derived, e.g., by using (\ref{jgen}):
\begin{equation}
\label{contract1}
 g_{\mu_{N_1} \sigma_{N_2}} \! \left\{ j_1, j_2, j_3 \right\}
= \left( n\!+\!2j_1\!+\!2j_2\!+\!j_3\!-\!1 \right)
  \left\{ j_1, j_2, j_3\!-\!1 \right\} +
  ( j_3\!+\!1)
  \left\{ j_1\!-\!1, j_2\!-\!1, j_3\!+\!1\right\} ,
\end{equation}
\begin{equation}
\label{contract2}
g_{\mu_{N_1\!-\!1} \mu_{N_1}} \left\{ j_1, j_2, j_3 \right\}
= \left( n\!+\!2j_1\!+\!2j_3\!-\!2 \right)
  \left\{ j_1\!-\!1, j_2, j_3 \right\}
  + 2\left( j_2\!+\!1 \right)
  \left\{ j_1, j_2\!+\!1, j_3\!-\!2 \right\}
\end{equation}
and an analogous formula for contraction with
$g_{\sigma_{N_2-1} \sigma_{N_2}}$. On the r.h.s.,
the structures $\{j_1,j_2,j_3\}$
should be set to zero whenever any of the $j$'s
becomes negative. Using these contractions, the following
recurrence relations for $\phi_{j_1 j_2 j_3; j'_1 j'_2 j'_3}$
can be obtained:
\begin{eqnarray}
\label{recur1}
N_1 N_2 \left\{ (n+N_1+N_2-j'_3-1) \;
\phi_{j_1 j_2 j_3; j'_1 j'_2 j'_3}
+ (j'_3-1) \;
\phi_{j_1 j_2 j_3; j'_1\!+\!1, j'_2\!+\!1, j'_3\!-\!2} \right\}
&& \nonumber \\
 = j_3 \; \phi_{j_1,j_2,j_3-1; j'_1, j'_2, j'_3-1}
& (j'_3>0) \, , &
\end{eqnarray}
\begin{eqnarray}
\label{recur2}
N_1 \; (N_1-1)
\left\{ (n+N_1+ j'_3-2) \; \phi_{j_1 j_2 j_3; j'_1 j'_2 j'_3}
+ 2 j'_2 \;\phi_{j_1 j_2 j_3; j'_1-1, j'_2-1,j'_3+2} \right\}
&& \nonumber \\
 = 2\;j_1 \; \phi_{j_1-1,j_2,j_3; j'_1-1,j'_2,j'_3}
& (j'_1>0) \, , &
\end{eqnarray}
and also a relation similar to (\ref{recur2}) but with interchanged
indices, $1\leftrightarrow 2$.
If either the second $\phi$ on the l.h.s.\ or the one on the
r.h.s.\ has a negative index, it is to be set to zero.

Let us consider some fixed values for $j_1$, $j_2$ and $j_3$ and suppose we
already know the $\phi$'s on the r.h.s.\ of, say, (\ref{recur1}),
corresponding to $(N_1-1)$ $\mu$'s and $(N_2-1)$ $\sigma$'s. The number of
equations (\ref{recur1}) is $T(N_1-1,N_2-1)$. If the number of tensor
structures does not increase between $(N_1-1,N_2-1)$ and $(N_1,N_2)$ (i.e.,
when $N_1$ and $N_2$ are odd), this means we have enough relations to
determine the $\phi$'s on the l.h.s.\ completely. On the other hand, if at
this step one extra tensor appears, we need one more relation between the
$\phi$'s. The same is true if we want to use (\ref{recur2}) to go from
$(N_1-2,N_2)$ to $(N_1,N_2)$. The additional information can be obtained by
contracting (\ref{decomp2}) with $k^{\mu_1} \ldots k^{\mu_{N_1}}
k^{\sigma_1} \ldots k^{\sigma_{N_2}}$ and using (cf. eq.~(B.10) of
\cite{DST})
\begin{equation}
I\left[ {(kp)}^{N_1} {(kq)}^{N_2} \right]
= \frac{ {(k^2)}^{(N_1+N_2)/2} }
       { 2^{(N_1+N_2)/2}
  {\left(\frac{n}{2}\right)}_{(N_1+N_2)/2} }
 \sum_{\{j\}} t_{j_1 j_2 j_3}
I\left[ (p^2)^{j_1} (q^2)^{j_2} (pq)^{j_3} \right] \, ,
\end{equation}
where $(a)_j \equiv \Gamma(a+j)/\Gamma(a)$ is the Pochhammer symbol.
Comparing the coefficients of
$I\left[ (p^2)^{j_1} (q^2)^{j_2} (pq)^{j_3} \right]$, we see that
\begin{equation}
\label{phi-c}
\sum_{\{j'\}}
\phi_{j_1 j_2 j_3; j'_1 j'_2 j'_3} \; t_{j'_1 j'_2 j'_3} =
\frac{1}{2^{(N_1+N_2)/2} \;
  {\left(\frac{n}{2}\right)}_{(N_1+N_2)/2} } \; .
\end{equation}

The recurrence relations (\ref{recur1})--(\ref{recur2}) can be solved
in several ways, leading to different representations for $\phi$. First,
we consider the special case when $j_1=0$, so that the
r.h.s.\ of eq.~(\ref{recur2}) vanishes and we get a relation
between $\phi_{0 j_2 j_3;j'_1 j'_2 j'_3}$ and
$\phi_{0 j_2 j_3;j'_1-1,j'_2-1,j'_3+2}$. Repeated application
of this relation yields
\begin{equation}
\label{phi023b}
\phi_{0 j_2 j_3;j'_1 j'_2 j'_3}
 = \frac{ {(-1)}^{j'_1} j'_2! }
        { j_2! \, {\left(\frac{n-2}{2}+j'_1+j'_3\right)}_{j'_1} }
\; \phi_{0 j_2 j_3;0 j_2 j_3} \; .
\end{equation}
To determine the normalization, we insert this result in
(\ref{phi-c}). The sum is a terminating $_2F_1$ series of unit
argument, which
can be written in terms of Pochhammer symbols. Substituting
the expression for $\phi_{0 j_2 j_3;0 j_2 j_3}$ thus found
back into (\ref{phi023b}) and using the symmetry~(\ref{sym:jj'})
gives
\begin{equation}
\label{phia023}
   \phi_{j_1 j_2 j_3; 0,(N_2-N_1)/2,N_1}
= \frac{ {(-1)}^{j_1} j_2! {\left(\frac{n-2}{2}\right)}_{j_1+j_3}   }
       { N_2! {\left(\frac{n}{2}\right)}_{(N_1+N_2)/2} {(n-2)}_{N_1} }
\; .
\end{equation}
To handle the general case, we move the second term in (\ref{recur2})
to the r.h.s. In this way, we get a recurrence relation in $j'_1$,
which has the following solution:
\begin{eqnarray}
\label{eq:phi-15-1}
   \phi_{j_1 j_2 j_3; j'_1 j'_2 j'_3}
&=& \frac{ j_1! j'_1! j'_2! }{ N_1! }
 \sum_{l=\mbox{\scriptsize max}\left(0,j'_1-j_1\right)
      }^{\mbox{\scriptsize min}\left(j'_1,j'_2\right)}
 \frac{ {(-1)}^l (j'_3+2l)! }
      { l! (j'_1-l)! (j'_2-l)! (j_1-j'_1+l)! }
\nonumber \\
&\times&
\frac{ {\left(\frac{n}{2}\right)}_{j'_3+2l} \;
       {\left(\frac{n-2}{2}\right)}_{j'_3+l} }
     { {\left(\frac{n-2}{2}\right)}_{j'_3+2l} \;
       {\left(\frac{n}{2}\right)}_{j'_1+j'_3+l} }
\;\; \phi_{j_1-j'_1+l,j_2,j_3; 0, j'_2-l, j'_3+2l}
\; .
\end{eqnarray}
We can write this as a familiar sum over $j$'s, (\ref{jsums}), by
defining $j''_1=j'_1-l$, $j''_2=j'_2-l$ and $j''_3=j'_3+2l$.
Substituting the result for $j'_1=0$ (\ref{phia023}),
we find
\begin{eqnarray}
\label{nice1}
   \phi_{j_1 j_2 j_3; j'_1 j'_2 j'_3}
&=&
\frac{j_1! j_2! j'_1! j'_2! {(-1)}^{(j_3-j'_3)/2} }{N_1! N_2!}
\sum_{\{j''\}}
\frac{ j''_3! }{ j''_1! j''_2!
 \left(\frac{j''_3-j_3}{2}\right)!
 \left(\frac{j''_3-j'_3}{2}\right)! }
\nonumber \\
&\times&
\frac{  {\left(\frac{n}{2}\right)}_{j''_3}
        {\left(\frac{n-2}{2}\right)}_{(j_3+j''_3)/2}
        {\left(\frac{n-2}{2}\right)}_{(j'_3+j''_3)/2} }
     {  {\left(\frac{n-2}{2}\right)}_{j''_3}
        {(n-2)}_{j''_3}
        {\left(\frac{n}{2}\right)}_{j''_1+j''_3}
        {\left(\frac{n}{2}\right)}_{j''_2+j''_3} } \; .
\end{eqnarray}
In (\ref{nice1}) and in the rest of this section, it is understood that
terms with factorials of negative integers in the denominator vanish. Note
that (\ref{nice1}) clearly shows the symmetries
$\phi_{j_1 j_2 j_3; j'_1 j'_2 j'_3}=
 \phi_{j'_1 j'_2 j'_3; j_1 j_2 j_3}=
 \phi_{j_2 j_1 j_3; j'_2 j'_1 j'_3}$.

Another special case for which we find a simple expression is when
$j_3=0$. This time, the r.h.s.\ of (\ref{recur1}) vanishes, and steps similar
to the ones described above lead us to
\begin{equation}
\label{phia120}
   \phi_{j_1 j_2 j_3; N_1/2,N_2/2,0}
= \frac{ \left(\frac{N_1}{2}\right)! \left(\frac{N_2}{2}\right)!
         {(-1)}^{j_3/2} j_3!
         {\left(\frac{n-1}{2}\right)}_{(N_1+N_2-j_3)/2}   }
       { N_1! N_2! 2^{j_3} \left(\frac{j_3}{2}\right)!
         {\left(\frac{n-1}{2}\right)}_{N_1/2}
         {\left(\frac{n-1}{2}\right)}_{N_2/2}
         {\left(\frac{n}{2}\right)}_{(N_1+N_2)/2}  } \; .
\end{equation}
Using (\ref{recur1}) as a recurrence relation in $j'_3$ gives
the following result:
\begin{eqnarray}
\label{eq:phi-14-3}
\!\!   \phi_{j_1 j_2 j_3; j'_1 j'_2 j'_3}
&=& \!\! \frac{ j_3! j'_3! }{ N_1! N_2! }
 \sum_{l=\mbox{\scriptsize max}\left(0,j_1-j'_1\right)}^{[j'_3/2]}
 \frac{ {(-1)}^l (2j'_1+2l)! (2j'_2+2l)! }
      { 4^l l! (j'_3-2l)! (j_3-j'_3+2l)! }
\nonumber \\
 &\times&
\frac{ {\left(\frac{n-1}{2}\right)}_{j'_1+j'_2+l} \;
       {\left(n\right)}_{2j'_1+2j'_2+4l} }
     { {\left(\frac{n-1}{2}\right)}_{j'_1+j'_2+2l} \;
       {\left(n\right)}_{2j'_1+2j'_2+j'_3+2l} }
\; \phi_{j_1,j_2,j_3-j'_3+2l; j'_1+l, j'_2+l,0}
\, .
\end{eqnarray}
By writing $j''_1=j'_1+l$, $j''_2=j'_2+l$ and $j''_3=j'_3-2l$,
and inserting (\ref{phia120})
we obtain a second representation for $\phi$:
\begin{eqnarray}
\label{nice2}
   \phi_{j_1 j_2 j_3; j'_1 j'_2 j'_3}
&=&  \frac{ j_3! j'_3! {(-1)}^{(j_3-j'_3)/2} 2^{j_1+j_2+j'_1+j'_2} }
          { N_1! N_2! }
\sum_{\{j''\}}
 \frac{ j''_1! j''_2! }
      { j''_3!
 \left(\frac{j_3-j''_3}{2}\right)!
 \left(\frac{j'_3-j''_3}{2}\right)! }
\nonumber \\
 &\times&
\frac{ {\left(\frac{n+1}{2}\right)}_{j''_1+j''_2} \;
       {\left(\frac{n-1}{2}\right)}_{(j_1+j_2+j''_1+j''_2)/2} \;
       {\left(\frac{n-1}{2}\right)}_{(j'_1+j'_2+j''_1+j''_2)/2} }
     { {\left(\frac{n-1}{2}\right)}_{j''_1+j''_2} \;
       {\left(\frac{n-1}{2}\right)}_{j''_1} \;
       {\left(\frac{n-1}{2}\right)}_{j''_2} \;
       {\left( n \right)}_{2j''_1+2j''_2+j''_3} }
\; .
\end{eqnarray}

The representations (\ref{nice1}) and (\ref{nice2})
complement each other. The number of non-vanishing terms in (\ref{nice1})
is $1+\mbox{min}(j_1,j_2,j'_1,j'_2)$, whereas the number in
(\ref{nice2}) is $1+\mbox{min}([j_3/2],[j'_3/2])$. Therefore,
depending on the values of the $j$'s and $j'$'s, it can be much
more efficient to use one or the other, even though both
lead to identical results\footnote{
It is possible to verify that (\ref{nice1}) reduces
to (\ref{phia120}) when $j'_3=0$, and (\ref{nice2}) to
(\ref{phia023}) when $j'_1=0$, with the help of a theorem on the
summation of $_5F_4$ series;
see \cite{Bailey}, {\S}4.3, eq.~3,
or \cite{Slater}, app.~III, eq.~13.}.
A simple example is when $N_1$ and $N_2$ are odd and $j'_3=1$.
In this case, (\ref{nice2}) consists of just one term, with
$j''_3=1$, which gives
\begin{equation}
\label{phia121}
   \phi_{j_1 j_2 j_3; (N_1-1)/2, (N_2-1)/2, 1}
= \frac{  \left(\frac{N_1-1}{2}\right)! \left(\frac{N_2-1}{2}\right)!
          {(-1)}^{(j_3-1)/2} j_3!
         {\left(\frac{n-1}{2}\right)}_{(N_1+N_2-j_3-1)/2}   }
       {  N_1! N_2! 2^{j_3}  \left(\frac{j_3-1}{2}\right)!
         {\left(\frac{n-1}{2}\right)}_{(N_1-1)/2}
         {\left(\frac{n-1}{2}\right)}_{(N_2-1)/2}
         {\left(\frac{n}{2}\right)}_{(N_1+N_2)/2}  } \; .
\end{equation}

An interesting feature of (\ref{nice1}) and (\ref{nice2}) is that, in each
term of the sum, the dependence on $j$ and on $j'$ is factorized. Both
formulae have the structure $\phi = M^T D M$, where $M$ is a triangular
matrix and $D$ is diagonal.
The matrix elements of the matrices $M$, and of their inverses
$M^{-1}$, can be presented as
\begin{equation}
\label{fact_of_life}
M_{jl} = \frac{(a)_{j+l}}{(l-j)!} \; b_l \;\;\; , \hspace{1cm}
(M^{-1})_{jl} = \frac{(-1)^{j-l} \; (a+1)_{2j}}
                     {(l-j)! \;\; (a)_{2j} \; (a+1)_{j+l}} \;
                \frac{1}{b_j} .
\end{equation}
With this decomposition, it is now easy to invert the
matrix $\phi$, which gives us the following
representations for the contraction matrix $\chi$:
\begin{equation}
\label{nicechi1}
\chi_{j_1 j_2 j_3; j'_1 j'_2 j'_3}
= \frac{{N_1}! {N_2}!}{{j_1}! {j_2}! {j'_1}! {j'_2}!} \;
\sum_{\{j''\}}
\frac{{j''_1}! {j''_2}!}{{j''_3}!
      \left( \frac{j_3-j''_3}{2} \right)!
      \left( \frac{j'_3-j''_3}{2} \right)!} \;
\frac{ (n-2)_{j''_3} \;
       \left(\frac{n}{2}\right)_{j''_3} \;
       \left(\frac{n}{2}\right)_{j''_1+j''_3} \;
       \left(\frac{n}{2}\right)_{j''_2+j''_3} }
     { \left(\frac{n-2}{2}\right)_{j''_3} \;
       \left(\frac{n}{2}\right)_{(j_3+j''_3)/2} \;
       \left(\frac{n}{2}\right)_{(j'_3+j''_3)/2} }
\end{equation}
if we use (\ref{nice1}), and
\begin{eqnarray}
\label{nicechi2}
\chi_{j_1 j_2 j_3; j'_1 j'_2 j'_3}
 &=& \frac{ N_1! N_2! }{ 2^{j_1+j_2+j'_1+j'_2} j_3! j'_3! }
\sum_{\{j''\}}
\frac{j''_3!}
     {j''_1! j''_2!
      \left( \frac{j''_3-j_3}{2} \right)!
      \left( \frac{j''_3-j'_3}{2} \right)! }
 \nonumber \\ & \times  &
\frac{ {\left(n\right)}_{2j''_1+2j''_2+j''_3}
       {\left(\frac{n-1}{2}\right)}_{j''_1}
       {\left(\frac{n-1}{2}\right)}_{j''_2}
       {\left(\frac{n+1}{2}\right)}_{j''_1+j''_2} }
     { {\left(\frac{n-1}{2}\right)}_{j''_1+j''_2}
       {\left(\frac{n+1}{2}\right)}_{(j_1+j_2+j''_1+j''_2)/2}
       {\left(\frac{n+1}{2}\right)}_{(j'_1+j'_2+j''_1+j''_2)/2}
     }
\end{eqnarray}
if we use (\ref{nice2}).
We also note that the determinant of $\phi$ is given
by the product of the diagonal elements of $M^T$, $D$ and $M$,
\begin{equation}
\label{detphi}
\det\phi = \prod_{\{j\}}
\frac{ {j_1}! \; {j_2}! \; {j_3}! \;
       {\left(\frac{n-2}{2}\right)}_{j_3} \;
       {\left(\frac{n}{2}\right)}_{j_3} }
     { {N_1}! \; {N_2}! \;
       {(n-2)}_{j_3} \;
       {\left(\frac{n}{2}\right)}_{j_1+j_3} \;
       {\left(\frac{n}{2}\right)}_{j_2+j_3} } .
\end{equation}

The matrices $M$ have a rather nice
interpretation. They transform the basis of tensor structures
$\{j_1,j_2,j_3\}$ into orthogonal bases. The orthogonal tensor
structures involved in our first representation, (\ref{nice1}) and
(\ref{nicechi1}), can be written as
\begin{eqnarray}
\label{orthonew}
\{\{ N_1, N_2 ; j_3 \}\}
 & \equiv &
2^{(N_1+N_2)/2} \; \sum_{\{j'\}}^{}
(-1)^{(j'_3-j_3)/2} \;
\frac{{j'_1}! {j'_2}!}{\left(\frac{j_3-j'_3}{2}\right)!} \;
{\textstyle\left(\frac{n-2}{2}\right)}_{(j_3+j'_3)/2} \;
\left\{ j'_1, j'_2, j'_3 \right\}
\nonumber \\
 & = &
\frac{ \partial^{N_1} }
 {\partial k_1^{\mu_1} \ldots \partial k_1^{\mu_{N_1}} } \;
\frac{ \partial^{N_2} }
 {\partial k_2^{\sigma_1} \ldots \partial k_2^{\sigma_{N_2}} } \;
(k_1^2)^{N_1/2} (k_2^2)^{N_2/2}
 C_{j_3}^{(n-2)/2}\left(\frac{(k_1k_2)}{\sqrt{k_1^2 k_2^2}}\right)
\nonumber \\
 & = &
\Box_{11}^{N_1/2} \, \Box_{22}^{N_2/2} \,
       C_{j_3}^{(n-2)/2}
\left(\frac{ \Box_{12} }
           {\sqrt{ \Box_{11}^{\phantom 2} \Box_{22}^{} }}\right)
{k_1}_{\mu_1} \ldots {k_1}_{\mu_{N_1}}
{k_2}_{\sigma_1} \ldots {k_2}_{\sigma_{N_2}} \, ,
\end{eqnarray}
where
\begin{equation}
\label{defGegenbauer}
C^{\gamma}_j (x) = \sum_{l=0}^{[j/2]}
 \frac{ {(-1)}^l {\left(\gamma\right)}_{j-l} }
      { l! (j-2l)! } \;
 {(2x)}^{j-2l}
\end{equation}
are Gegenbauer polynomials. If we remember that $N_1=2j_1+j_3$ and
$N_2=2j_2+j_3$, we see that (\ref{orthonew}) does not really contain
negative powers or square roots of d'Alembertians. The orthogonality of the
structures (\ref{orthonew}) can be explained using the fact \cite{GPXT}
that
\begin{equation}
\Box_{11}
(k_1^2)^{j_3/2} (k_2^2)^{j_3/2}
 C_{j_3}^{(n-2)/2}\left(\frac{(k_1k_2)}{\sqrt{k_1^2 k_2^2}}\right)
 = 0
\end{equation}
(and the same for $\Box_{22}$),
and it is closely related to the orthogonality of the Gegenbauer
polynomials with respect to integration over the unit sphere
in $n$-dimensional Euclidean space.

An analogous explanation can be given for our second representation,
(\ref{nice2}) and (\ref{nicechi2}). The corresponding orthogonal
tensor structures are connected with polynomials that are annihilated
by $\Box_{12}$,
\begin{equation}
\label{R-polynomial}
(k_1^2)^{j_1} \; (k_2^2)^{j_2} \;
 \sum_{l=0}^{\mbox{\scriptsize min}\left(j_1,j_2\right)} \;
 \frac{ {(-1)}^l \; \left( \frac{n-1}{2} \right)_{j_1+j_2-l} }
      { l! \; (j_1-l)! \; (j_2-l)! } \;
\left( \frac{(k_1 k_2)^2}{k_1^2 k_2^2} \right)^l \; .
\end{equation}

For completeness, we mention that a third representation for
$\phi_{j_1 j_2 j_3; j'_1 j'_2 j'_3}$ can be derived by using (\ref{recur2})
in the opposite direction, as a recurrence relation in
$j'_3$. It is a sum of $1+\mbox{min}(j_1,[j'_3/2])$ terms, in which
the dependence on $j$ and $j'$ is factorized. However,
it is less symmetric than (\ref{nice1}) and (\ref{nice2}) and does
not appear to have such a simple interpretation.
It is also possible to write (\ref{recur1}) as a recurrence relation
in $j'_1$, but it is more difficult to solve than the other ones.

To conclude this section, let us insert our first expression for
$\phi$ (\ref{nice1}) into the reduction formula (\ref{decomp2}).
The result can be written in the following way\footnote{
In eq.(6) of ref.~\cite{Tarasov-Pisa}, the scalar coefficients are also
written in terms of Gegenbauer polynomials, but a different basis is used
for the tensor structures. By some transformations, and the summation
of a hypergeometric series of unit argument, one can show that the result
is equivalent to (\ref{I-Geg}).}:
\begin{eqnarray}
\label{I-Geg}
\lefteqn{\hspace*{-23mm}
I\left[ p_{\mu_1} \ldots p_{\mu_{N_1}} \,
         q_{\sigma_1} \ldots q_{\sigma_{N_2}} \right]
 = \frac{ 1 }{ 2^{N_1+N_2} }
\sum_{\{j\}}
\frac{ j_3! }{ j_1! j_2!}
\frac{  {\left(\frac{n}{2}\right)}_{j_3}     }
     {  {\left(\frac{n-2}{2}\right)}_{j_3}
        {(n-2)}_{j_3}
        {\left(\frac{n}{2}\right)}_{j_1+j_3}
        {\left(\frac{n}{2}\right)}_{j_2+j_3} }
} && \nonumber \\ & \times &
 I\left[(p^2)^{N_1/2} (q^2)^{N_2/2}
       C_{j_3}^{(n-2)/2}\left(\frac{(pq)}{\sqrt{p^2 q^2}}\right)\right]
\{\{ N_1, N_2 ; j_3 \}\}
 \, .
\end{eqnarray}
When contracted with $k_1$ and $k_2$, this gives
\begin{eqnarray}
\lefteqn{\hspace*{-5mm}
I\left[ {\left(k_1p\right)}^{N_1}
        {\left(k_2q\right)}^{N_2} \right]
= \frac{ N_1! N_2! }{ 2^{N_1+N_2} }
\sum_{\{j\}}
\frac{ j_3! }{ j_1! j_2!}
\frac{  {\left(\frac{n}{2}\right)}_{j_3}     }
     {  {\left(\frac{n-2}{2}\right)}_{j_3}
        {(n-2)}_{j_3}
        {\left(\frac{n}{2}\right)}_{j_1+j_3}
        {\left(\frac{n}{2}\right)}_{j_2+j_3} }
} &&\nonumber \\
& \times & \!\!\!
 I\left[(p^2)^{N_1/2} (q^2)^{N_2/2}
       C_{j_3}^{(n-2)/2}\left(\frac{(pq)}{\sqrt{p^2 q^2}}\right)\right]
 (k_1^2)^{N_1/2} (k_2^2)^{N_2/2}
 C_{j_3}^{(n-2)/2}\left(\frac{(k_1k_2)}{\sqrt{k_1^2 k_2^2}}\right) \, ,
\end{eqnarray}
and, similarly, the three-point functions (\ref{eq:3point}) can
be expressed as
\begin{eqnarray}
\label{C-Geg}
C(k_1,k_2) & = &
\sum_{N_1,N_2}  \frac{ 1 }{ 2^{N_1+N_2} }
\sum_{\{j\}}
\frac{ j_3! }{ j_1! j_2!}
\frac{  {\left(\frac{n}{2}\right)}_{j_3}     }
     {  {\left(\frac{n-2}{2}\right)}_{j_3}
        {(n-2)}_{j_3}
        {\left(\frac{n}{2}\right)}_{j_1+j_3}
        {\left(\frac{n}{2}\right)}_{j_2+j_3} }
\nonumber \\
  &  \times &
 (k_1^2)^{N_1/2} (k_2^2)^{N_2/2}
 C_{j_3}^{(n-2)/2}\left(\frac{(k_1k_2)}{\sqrt{k_1^2 k_2^2}}\right)
\nonumber \\
  &  \times &
\left\{
\Box_{11}^{N_1/2} \, \Box_{22}^{N_2/2} \,
       C_{j_3}^{(n-2)/2}
\left(\frac{ \Box_{12} }
           {\sqrt{ \Box_{11}^{\phantom 2} \Box_{22}^{} }}\right)
C(k_1,k_2)
\right\}_{k_1=k_2=0} \; .
\end{eqnarray}


\newpage

{\bf 4. Conclusion}

The main results of this paper are the explicit formulae (\ref{nice1}) and
(\ref{nice2}) for the coefficients $\phi_{j_1 j_2 j_3; j'_1 j'_2 j'_3}$,
which are necessary for both the reduction of two-loop vacuum tensor
integrals of arbitrary rank to scalar integrals (\ref{decomp2}), and for
the construction of the projectors (\ref{projectors}) that extract the
Taylor coefficients of scalar three-point diagrams. The
tensor integrals are written in a more compact way (\ref{I-Geg}) by using
orthogonal bases involving Gegenbauer polynomials for the tensor structures
and scalar integrals. A similar representation is also derived for the
three-point functions, eq.~(\ref{C-Geg}). This shows, once again, that when
one expands in $(k_1 k_2)$, the basis of the Gegenbauer polynomials is
useful \cite{GPXT} (see also in \cite{Tarasov-Pisa}). On the way, we have
also obtained explicit expressions for the elements of the contraction
matrix $\chi$, (\ref{nicechi1}) and (\ref{nicechi2}).

The elements of the decomposition matrix $\phi$ are
expressed as single finite sums of quotients of Pochhammer symbols
involving the space-time dimension $n$. Whenever one of the indices $j$ or
$j'$ reaches its minimal value (i.e. $j_1=0$, $j_2=0$, $j_3=0$ or $j_3=1$),
we have just a single term (eqs.~(\ref{phia023}), (\ref{phia120}) and
(\ref{phia121})). In most cases, however, a sum of several terms cannot be
avoided because the numerator of $\phi_{j_1 j_2 j_3;j'_1 j'_2 j'_3}$
contains non-factorizable quadratic (or higher degree) factors, e.g.:
\begin{equation}
\phi_{1,1,2;1,1,2} = \frac{n^2+3n+6}
  { 72 (n - 1) n (n + 1) (n + 2) (n + 4) (n + 6) } \; .
\end{equation}
Therefore, it seems unlikely that more simple general formulae for
$\phi_{j_1 j_2 j_3;j'_1 j'_2 j'_3}$ than (\ref{nice1}) and (\ref{nice2})
can be found.

Numerous checks on the results were performed by computer algebra
\cite{comp}. We also verified, for a large number of cases, that the
reduction formula in ref.~\cite{Tarasov-Pisa} is equivalent to ours. The
reduction algorithm in \cite{Ch-AI93} is of a different kind, because it is
based on integration by parts \cite{ibp} and depends on the particular form
of the propagators in the diagrams.

Let us now briefly discuss the application of these formulae. As an
example, we consider the momentum expansion of a massive two-loop
three-point function (see e.g. in \cite{FT,Tarasov-Pisa}). Here, one can
choose between two methods. The first is to expand all propagators in
powers of the external momenta, yielding a collection of vacuum integrals
with tensor numerators. Then, all the tensor integrals are reduced to
scalar integrals times invariant tensor structures $\{j_1,j_2,j_3\}$. After
that, the integrals can be calculated by expressing the numerators in terms
of inverse propagators and cancelling them against the denominators. The
final step is to contract the tensor structures with the external vectors
(momenta or polarization vectors), which gives us a polynomial in scalar
products of those vectors. If two or more of the external vectors are
identical, each term is accompanied by a symmetry factor which, in general,
can be written as a multiple sum of combinations of factorials, or computed
by a recursive algorithm.

The second method is to decompose the three-point function into scalar
form factors, which are then expanded in the external invariants by using
d'Alembertian operators. It takes some work to apply the d'Alembertians
to the integrand, but in return, we no longer need to contract any tensor
structures with external vectors. After setting the external momenta to
zero, we get scalar vacuum integrals, which are evaluated in the same way as
before. For given $N_1$ and $N_2$, we end up with three sums: one over
powers of the external momenta, one over powers of the d'Alembertians, and
the internal sum in the representation of $\phi$. As we have already
mentioned, two of these sums can be absorbed in Gegenbauer polynomials, in
which case only one extra sum remains.

\vspace{5mm}
{\bf Acknowledgements.}
We thank K.G.~Chetyrkin and O.V.~Tarasov for helpful discussions. We wish
to express our gratitude to the Aspen Center for Physics, where an
essential part of this work was done, for their hospitality, and to
the US Department of Energy for supporting A.D.'s visit to Aspen. A.D.'s
research was supported by the Research Council of Norway, and it was also
related to the research programme of the INTAS--93-0744 project. J.B.T. was
supported by the Graduiertenkolleg ``Teilchenphysik'' in Mainz.

\vspace{3mm}

\end{document}